# Taking *Saratoga* from Space-Based Ground Sensors to Ground-Based Space Sensors

Lloyd Wood
Centre for Communication Systems Research
University of Surrey
Guildford GU2 7XH
England, United Kingdom
L.Wood@surrey.ac.uk

Charles Smith
Commonwealth Scientific and Industrial Research Organisation
Cnr., Vimiera and Pembroke Roads
Marsfield, New South Wales 2122
Australia
charles.smith@csiro.au

Wesley M. Eddy
MTI Systems
NASA Glenn Research Center
21000 Brookpark Road
Cleveland, OH 44135
United States
wes@mti-systems.com

Will Ivancic
NASA Glenn Research Center
21000 Brookpark Road,
Cleveland, OH 44135
United States
William.D.Ivancic@grc.nasa.gov

Chris Jackson
Surrey Satellite Technology Ltd
Tycho House, 20 Stephenson Road
Surrey Research Park, Guildford GU2 7YE
England, United Kingdom
C.Jackson@sstl.co.uk

*Abstract*—The *Saratoga* transfer protocol was developed by Surrey Satellite Technology Ltd (SSTL) for its Disaster Monitoring Constellation (DMC) satellites. In over seven years of operation, *Saratoga* has provided efficient delivery of remote-sensing Earth observation imagery, across private wireless links, from these seven low-orbit satellites to ground stations, using the Internet Protocol (IP). *Saratoga* is designed to cope with high bandwidth-delay products, constrained acknowledgement channels, and high loss while streaming or delivering extremely large files. An implementation of this protocol has now been developed at the Australian Commonwealth Scientific and Industrial Research Organisation (CSIRO) for wider use and testing. This is intended to prototype delivery of data across dedicated astronomy radio telescope networks on the ground, where networked sensors in Very Long Baseline Interferometer (VLBI) instruments generate large amounts of data for processing and can send that data across private IP- and Ethernet-based links at very high rates. We describe this new *Saratoga* implementation, its features and focus on high throughput and link utilization, and lessons learned in developing this protocol for sensor-network applications.

### TABLE OF CONTENTS

## 1. INTRODUCTION

Private computer networks can have very different operating paradigms and underlying assumptions from that of the public Internet. In the public Internet, congestion of resources (i.e. router queues and available link capacity) is caused by competition between unsynchronized applications running on end hosts with different owners with different goals. In private networks, tools for flow management and traffic engineering are available within an autonomous system under single management. Often, the primary requirement of a private network is simply to support moving data from A to B as quickly as possible, to allow a task that requires that data to proceed. In a network where all nodes and end hosts are owned, operated and managed by a single entity, network congestion due to competition may not be a concern. Coarse-grained scheduling across time of separate individual data transfers, in sequence one after another, can avoid competition, allowing each data transfer and the overall series of transfers to be completed as quickly as possible without devoting time to inefficient competition for resources or congestion control loops.

Such a scenario is present in copying image data from low-Earth-orbiting remote-sensing satellites to ground stations during overhead passes lasting less than fourteen minutes' duration. As much data must be transferred in this time as possible, in order to make the most use of the available downlink and of the satellite capabilities. This data should be carried as quickly as each satellite downlink permits.

When the remote-sensing satellite communications are built around reliance on the Internet Protocol (IP), a fast IP-based transport protocol becomes necessary to deliver the image data. The popular Transmission Control Protocol (TCP), which is used across the Internet, includes algorithms such as slow-start and congestion avoidance, which attempt to sense network capacity limits and remain well below the available capacity to ensure fairness between flows. TCP assumes that any lost packet indicates congestion and that backoff is needed. TCP reduces its sending rate accordingly. When losses are solely due to channel errors, TCP's assumptions no longer hold, and its reaction prevents efficient link utilization. A different transport protocol, holding different assumptions about its operating environment, can safely improve performance in this scenario.

## 2. CREATION OF *SARATOGA*

Surrey Satellite Technology Ltd (SSTL) uses IP for payload communications on its Disaster Monitoring Constellation (DMC) satellites, and has also transitioned from AX.25 to IP for platform Telemetry, Tracking and Command (TT&C). This IP use is built upon earlier experiments done with uploading an IP stack to an onboard computer on the earlier UoSAT-12, with Hogie *et al*. [1]. Integration with the terrestrial Internet, with use of cheap commercially-available routing equipment and personal computers in ground station local area networks (LANs), is a benefit of this approach. A number of demonstrations of integration with the terrestrial Internet and remote operations have been undertaken [2].

As of this writing, seven DMC satellites have been launched to orbit since 2002, of which two (AlSAT-1, launched 2002, and BilSAT, launched 2003) have now completed their missions and reached end of operational life due to onboard batteries no longer retaining their charges [3][4]. All DMC satellites use IP to transfer raw Earth imaging sensor data, at downlink rates from an initial 8.1 Mbps (coincidentally the maximum rate of the serial interface on the Cisco routers introduced to SSTL by Hogie) to 20/40/80 Mbps on more recent DMC satellites. New DMC launches are planned, with 105/210 Mbps downlinks for these missions.

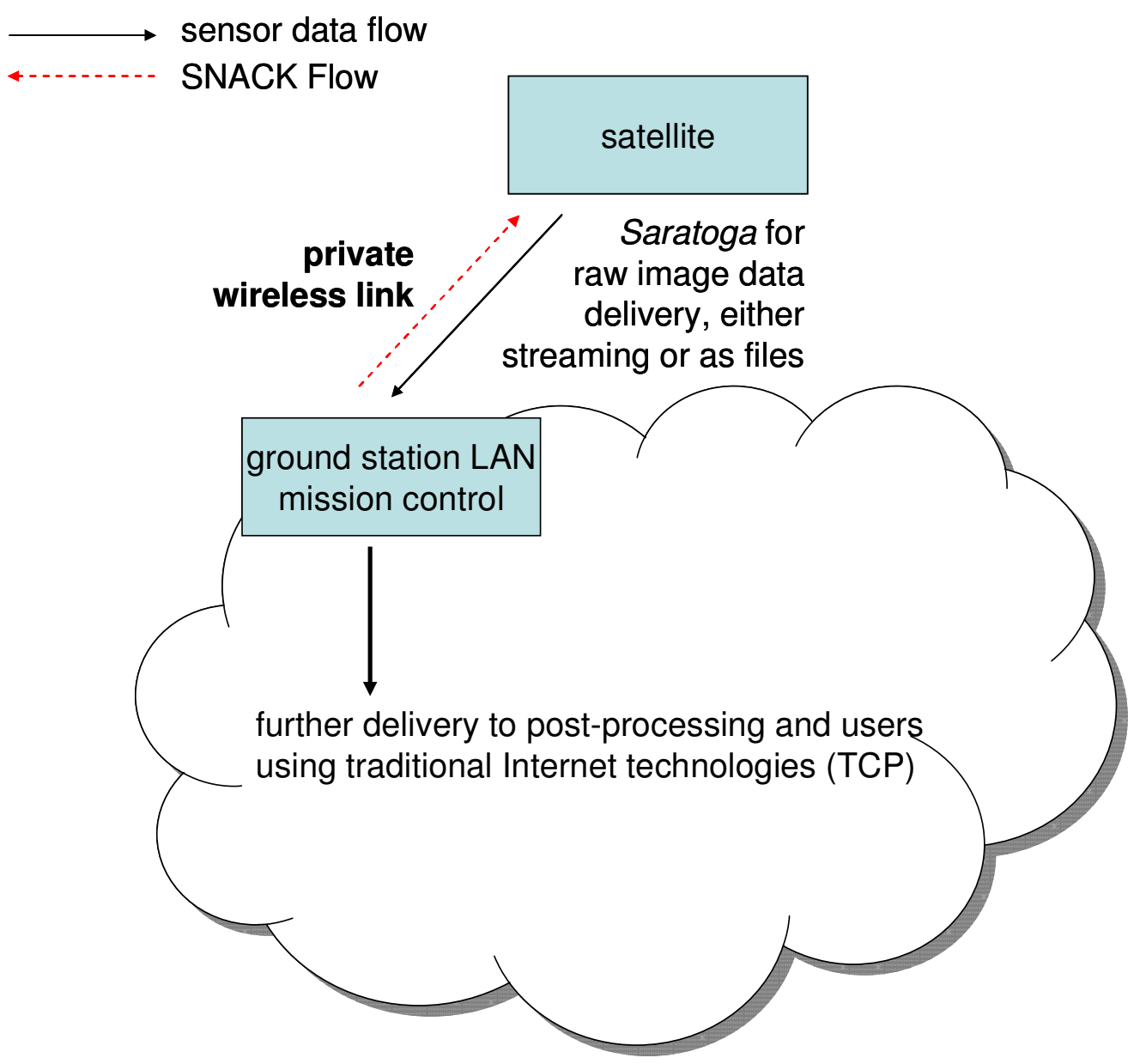

**Figure 1 – Use of *Saratoga* for remote-sensing satellites**

SSTL initially downloaded imagery over IP and these downlinks by using the first in-space deployment of the Consultative Committee for Space Data Systems File Delivery Protocol (CCSDS CFDP), onboard AlSAT-1 [5]. SSTL then developed its own replacement Internet transport protocol in-house, to increase performance and transfer data as quickly as the available downlink capacity and low-end PowerPC processor capability onboard would permit. This was named *Saratoga*, for the USS Saratoga, sunk in the Pacific near Bikini Atoll (which the protocol's designer, Jackson, has dived). The *Saratoga* protocol design has been described and enhanced over time, and a recent version of the protocol has been specified in detail to the Internet Engineering Task Force [6].

*Saratoga* adds selective negative acknowledgements (SNACKs) above the well-known User Datagram Protocol (UDP), enabling reliable delivery of files via retransmissions when packets are corrupted and lost due to channel errors, but without the assumptions about congestion that the file transfer protocol (FTP) running over TCP inherits. *Saratoga*'s use in delivering raw image data from satellites, complementing use of more familiar Internet technologies on the ground, is illustrated in Fig. 1.

*Saratoga* has also enabled delivering data as large 'bundles' for the first in-space tests of the 'Bundle Protocol' and 'Interplanetary Internet' from the UK-DMC satellite [7]. Acting as a 'bundle convergence layer' was proposed as an optional feature for *Saratoga* for delay-tolerant networking scenarios where the Bundle Protocol might be used [8].

## 3. FEATURES OF *SARATOGA*

As well as being designed to run as fast as possible to fill a link, *Saratoga* has a number of useful features:

- File advertisement, requests with directory browsing, and reliable delivery of files, with strong end-to-end checksums if desired.

- Streaming. The ability to send continuous data at high rates in real time, either reliably or unreliably.

- The ability to scale to deliver extremely large files or fast streams, if required. This was motivated by the observation that imaging files being created onboard early DMC satellites were already hundreds of megabytes or gigabytes in size. Saratoga scales across multiple implementation environments by supporting either very large or relatively smaller limits on file sizes. 4GiB (gibibytes) is a threshold; below it the position in a file can be described with 32-bit offsets where each bit increment represents a byte, while above that size 64-bit offsets are needed. Also supporting 16-bit offset fields for transferring very small files (up to 64KiB) and 128-bit offsets for very large files (16,384 pebibytes or above in size) makes *Saratoga* able to scale up or down and future-proof across any conceivable file or stream size – although individual implementations do not need to take advantage of all of these offset sizes, and can restrict themselves to using and advertising support for a subset of these sizes. Low-end devices with eight-bit processors might only ever support and send small files using 16-bit offsets, for example. Files created and being delivered for astronomy needs are unlikely to need more than 64 bits to describe file size or offset position within the file… any time soon.

- Support for link-local multicast, to send to multiple receivers simultaneously and efficiently. This can enable simultaneous software uploads to multiple devices.

- Functionality in constrained asymmetric environments, where there is a heavily restricted backchannel for acknowledgements to the data flow in the forward path. On the DMC satellites, uplinks are typically below 38.4kbps to support downlinks over 850 times faster. This is less of a concern for radio astronomy, where fibre can be utilized in both directions, but efficiency in the control channel can decrease fibre deployment, as is discussed later.

- Use of UDP, which allows ease of implementation on computers in application 'userland' rather than in kernel space, with applications working off established port numbers, and eases working through network address translation (NATs) and firewalls, and with multicast, for longer-distance communication along multi-hop paths if required. Although *Saratoga*'s use is envisaged as primarily across single hops rather than across longer paths, these advantages were considered to be worth the use and small framing overhead cost of the UDP header, which avoided 'reinventing the wheel.' With line-rate UDP drivers available under most operating environment implementations, and the SNACK mechanism providing reliable delivery over a UDP transport, this can maximize link utilization.

- Optional UDP-Lite support for data delivery can allow delivery of data corrupted in transit, if an application is able to cope with and detect errors. This can be preferable in some scenarios to discarding entire packets, which turns errored bits into erased frames. Delivering packets with errored payloads is rarely useful, but in practice, when coupled with a strong layer-2 frame cyclic redundancy check (CRC), UDP-Lite minimizes the amount of payload checksumming required, and is preferable to turning off UDP checksums entirely as vital headers are still checked.

Fig. 2 compares *Saratoga* to an equivalent TCP flow. While it would be possible to modify a TCP implementation to remove the slow start algorithm and change other congestion-related behaviour, including increasing buffer sizes, TCP remains buffer and window-limited. Matching TCP to the bandwidth-delay product for a thousand-kilometre 100Gbps link, leading to a 40MB send buffer space, is unusual for TCP implementations. SSTL's DMC satellites do not use TCP in their onboard computers.

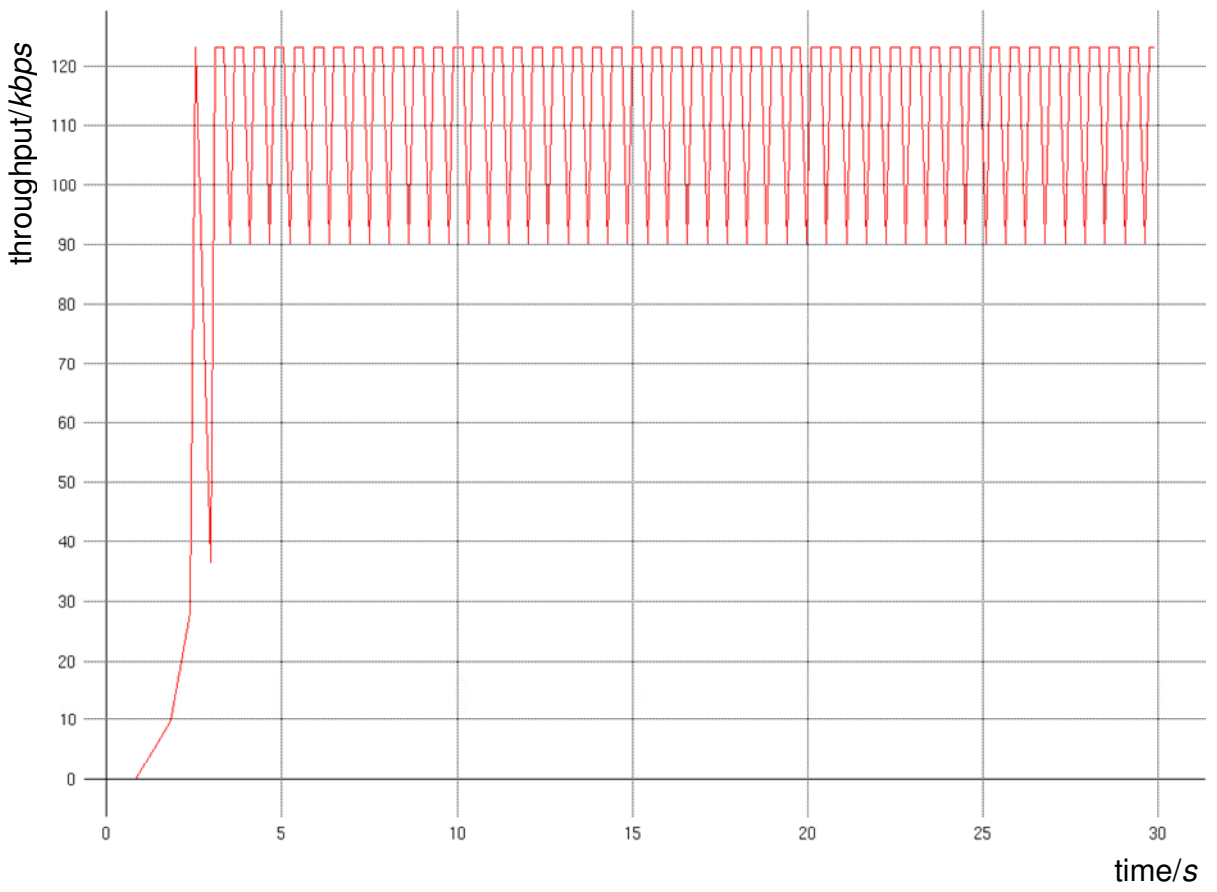

a. *ns2* [9] simulation of a single FTP flow across a reliable long-distance 128kbps link. TCP SACK repeatedly probes available capacity and increases its rate to above the link rate, causing the output queue to drop packets, which leads to backoff.

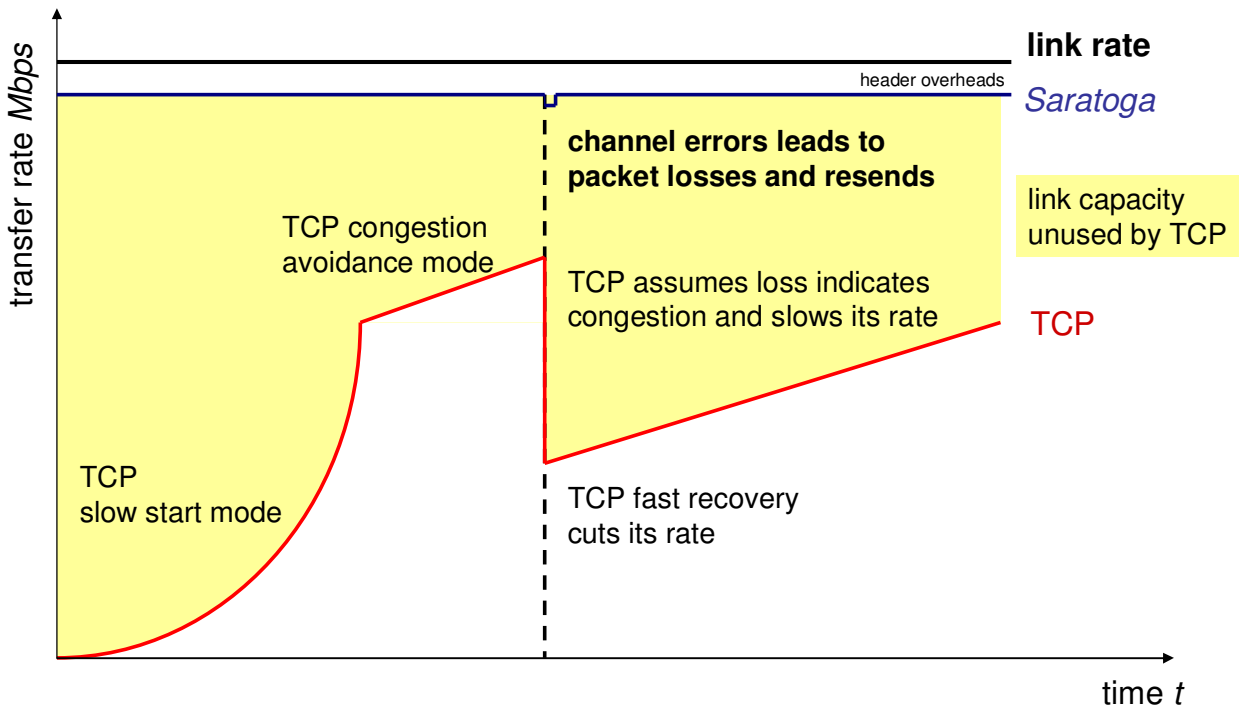

b. Identification of various TCP behaviours.

**Figure 2 –*Saratoga* and TCP reacting to packet loss**

TCP's head-of-line blocking can also make it unsuitable for real-time streaming, as anyone who has played stuttering video clips on the web will recognize.

## 4. Radio Astronomy Networks

A number of distributed radio astronomy installations, where large amounts of digital data must be generated, moved and stored, are under construction or being proposed. A number of these are being constructed as pathfinders to gain experience for design and construction of the Square Kilometre Array (SKA) [10].

The construction costs of these radio telescopes are to a great extent determined by the deployment costs of the fibre optic networks needed to transport data from sensors to processors [11]. Improved distributed radio telescopes such as the Atacama Large Millimeter Array (ALMA), the e-MERLIN fibre optic upgrade to the microwave-using Multi-Element Radio Linked Interferometer Network (MERLIN) and the Expanded Very Large Array (eVLA), would not have been possible without and are dependent on optical fibre technology [12]. Fibre optic data transport infrastructure is a critical requirement for emerging sensor technologies, including high-density and low-density aperture arrays and phased-array feeds.

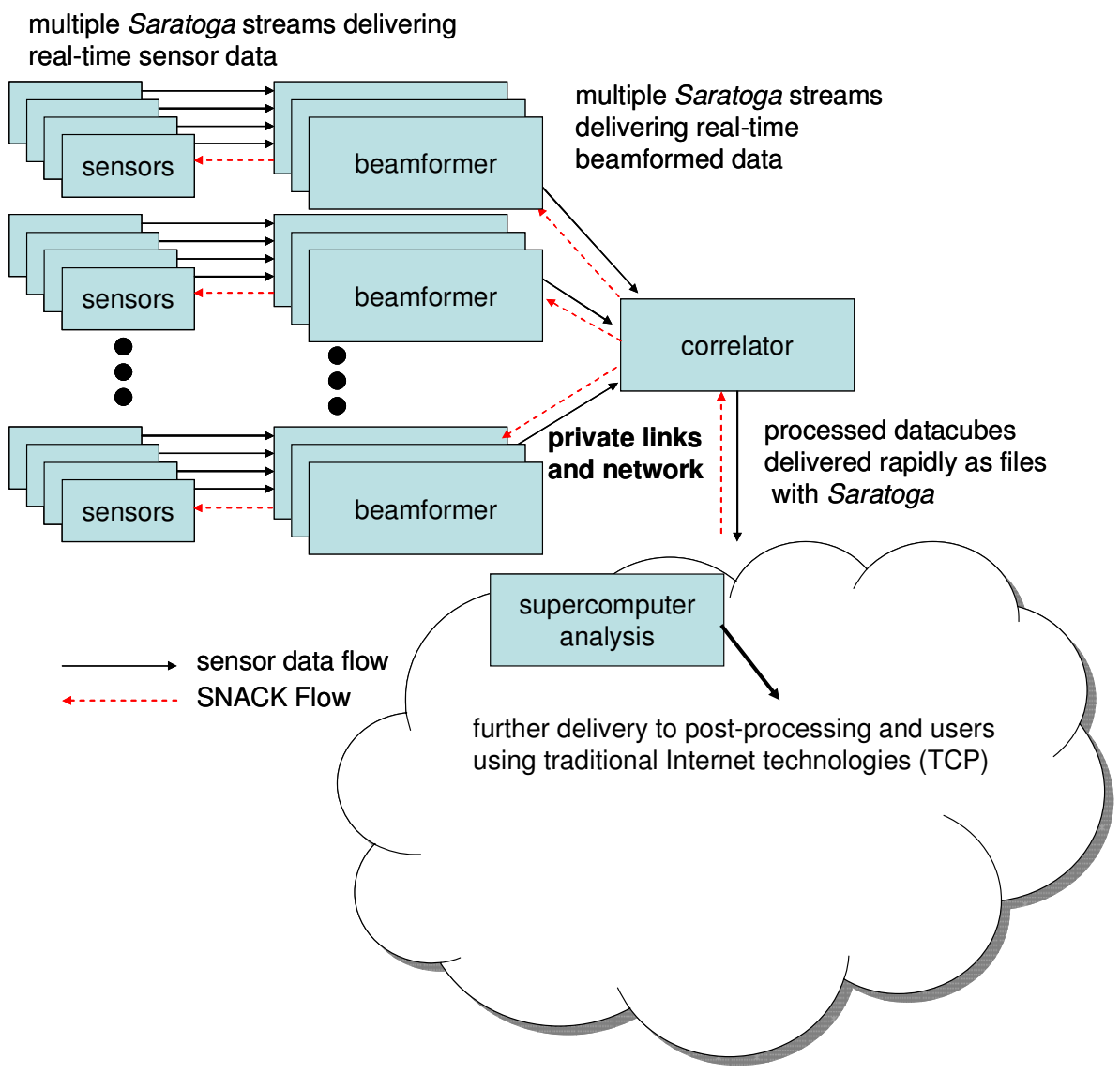

**Figure 3 – proposed uses of *Saratoga* in radio astronomy**

These new technologies increase the data output requirements from existing sensors using single-pixel feeds by up to two orders of magnitude.

The phased-array feeds being developed by CSIRO in Australia for the Australian Square Kilometre Array Pathfinder Telescope (ASKAP) are expected to stream 192 parallel 10 Gbps feeds from each of the 36 twelve-metre dish receivers. This gives a total of just under 70 Tbps, or about eight terabytes per second. The ASKAP project is a 1% pathfinder demonstrator for the planned SKA radio telescope, which is expected to have 80% of its receivers located at a central site in either Western Australia or the Karoo in South Africa, with the remaining 20% of receivers spread across thousands of kilometres in stations across either Australia & New Zealand or across Southern Africa. The decision on the final SKA site is expected to be made in 2012.

Optical fibre interconnects are critical, both at the central site and from remote stations to a single correlation facility given the immense sensor data payloads [13]. It is desirable to be able to minimize design and construction costs by using commercially-available equipment where possible, to exploit Moore's law and available commercial products [14].

Future radio astronomy networks are currently being designed in anticipation of where commercial equipment will be in several years, once procurement and construction have begun. For example, 100 Gbps long-range optical fibre Ethernet links, or better, can be expected to be specified for networking use and the construction phase of the SKA project, rather than being limited to current state of the art.

The drive to be able to use commercial networking equipment, and avoid spending money on developing custom solutions where possible, is the same underlying motivation as in adopting Internet technologies in the DMC satellites. However, just like the processing performance required, the performance of the networking technology needed for radio astronomy will lie at the high, rather than the low, end of possible requirements. Engineering costs can still be reduced by leveraging the capabilities of existing commercial optical fibre, Ethernet and IP-based networking technologies. In this context – high-speed private networks supporting data delivery for radio astronomy – there is still a need to be able to use the available link capacity as efficiently as possible. A single TCP flow or few TCP flows cannot fill a 100 Gbps fibre link efficiently or rapidly, due to TCP's assumptions and resulting behaviour.

We believe that *Saratoga*'s streaming facility will be useful for sending real-time data back from individual distributed sensors. The raw sensor data is beam-formed on-site for an initial reduction to meet the link capacity requirements, and then streamed at a fixed rate to a central correlator for processing as outlined in Fig. 3.

The data flow in radio astronomy sensor networks is inherently asymmetric, flowing from the sensors. The sensing, beam-forming and correlation tasks do not require a bidirectional exchange between the different stages of computation. With the advent of the new array receiver technologies, a unidirectional link capability is most desirable, as it immediately reduces the requirements for fibre, transmitters and receivers by half. However, eliminating any form of feedback between the computation stages leads to added software complexity in order to ensure the validity and robustness of the data stream. With its SNACK capability, *Saratoga* provides a necessary mechanism to monitor and guarantee the validity of data delivery, while minimizing the return path data flow.

An example can be for a focal plane array of 200 sensors, each transmitting at 10 Gbps to a beam-former computation engine, with a single 10 Gbps return path providing the multiplexed error return and acknowledgement capability for all 200 sensors. At the post-processing stage with *Saratoga's* support for the inclusion of extremely accurate timestamps on each data frame, the timing accuracy required for the beam-forming and correlation tasks can be captured and maintained throughout the computation phases without the need for duplicating timing structures within each data frame.

Optical fibre is now a relatively noise-free medium, but with non-zero error rates of typically 1 in $10^{-9}$ or $10^{-12}$ or better with inline amplification on extremely long links, a corrupted and discarded frame can be expected for roughly every gigabit of data transferred. This is compensated for by SNACKs and resends. *Saratoga*'s inherent ability to efficiently transport and guarantee error-free delivery of extremely large files with its flexible offset size will also be useful for passing post-processed image 'data cubes' around for later analysis across high-performance links. These two possible applications for *Saratoga* are shown in Fig. 3.

An implementation of *Saratoga*, with support for streaming raw data, and for file delivery, has been developed at CSIRO. This is intended to anticipate and meet radio astronomy network needs, and could be used for data delivery in the Square Kilometre Array. Performance testing of *Saratoga* can be undertaken over dedicated 10 Gbps and 40 Gbps optical circuits across a 1000 km span in Australia.

It has been calculated that, to transmit streaming data directly from the 10 Gbps focal plane array sensors, reaching a minimum link utilization of 87% is required to carry a 12-bit sampled stream, and if 95% link utilization is attained, the sample size can be increased to 13 bits, even after the necessary overheads for Ethernet jumbo frames, and IP, UDP and *Saratoga* header overheads.

An example calculation of link utilization for a simple sensor scenario is in Appendix A.

## 5. SENSOR STREAMS AND IMAGE DATA CUBES

The output stream from a correlator is processed by a supercomputer to generate a multi-dimensional image data cube, which is then further processed and analysed by radio astronomers to develop and test their research hypotheses.

An image data cube is a three-dimensional representation of the sky, where the *x* axis holds an index to the declination angle (Dec), the *y* axis holds an index to the right ascension angle (RA) and the *z* axis holds an index to the cosmological red shift (Z). This is illustrated in Fig. 4.

(Although this is traditionally called a cube, the sides are rarely of equal length. We believe that *data brick* is a more accurate term.)

To give an idea of the scale of data produced by these radio astronomy arrays, let's consider a couple of examples:

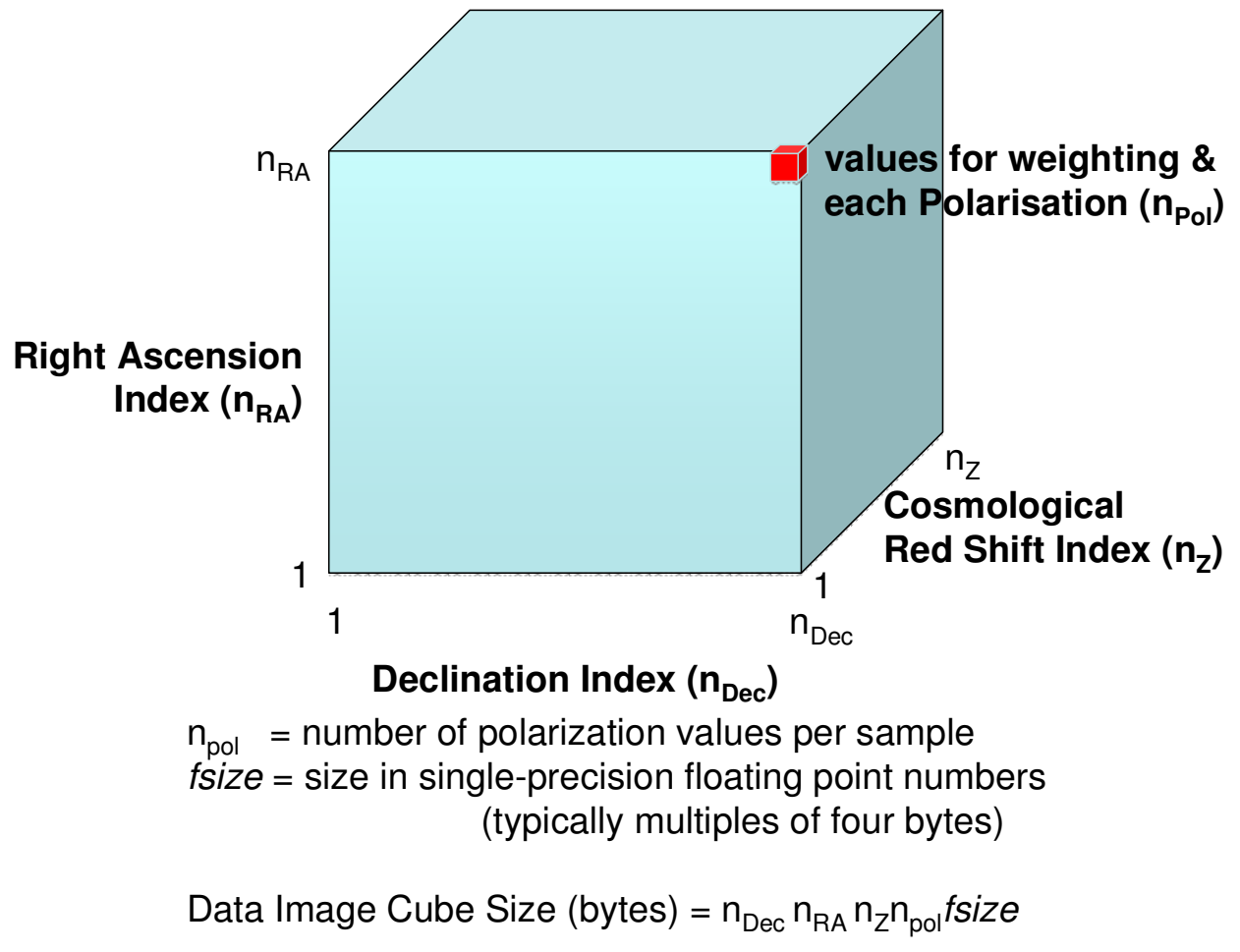

**Figure 4 – Image Data Cube Representation**

*Murchison Widefield Array*

The Murchison Widefield Array (MWA), at the Murchison Radio Observatory in Western Australia, consists of 8,192 dual-polarization dipole antennas intended for sensing the 80-300 MHz frequency range [15].

These are arranged as 512 "tiles," each being a 4 x 4 array of dipoles. An image data cube is generated every twelve minutes, with 2,700 $n_{RA}$ by 2,700 $n_{Dec}$ by 768 $n_Z$. Each indexed point in the cube holds a single-precision floating point (4 bytes) weight and four single-precision floating-point polarizations (16 bytes $n_{Pol}$), for a total of 20 bytes per point.

2,700 x 2,700 x 768 x 20 bytes yields a 112 Gigabyte image data cube that is generated every twelve minutes during an observation period and must be transmitted. (As the cube is transmitted, we use SI decimal rather than IEC binary prefixes of magnitude, to be consistent with the convention for communications equipment). Delivery of these cubes as files can be thought of as equivalent to a continuous stream of 1.25 Gbps – but remember that that is only for post-processed data cubes without transport overheads, and not for the raw sensor data, which must be streamed at a much higher overall rate, including network encapsulation overheads. The data rate streaming from the correlator is estimated as 19Gbps [16].

*Australian Square Kilometre Array Pathfinder (ASKAP)*

The ASKAP telescope, currently under construction at the Murchison Observatory site, is planned to consist of 36 12-metre dishes with each dish holding 192 phased-array feed sensors (that is, 96 dual-polarisation sensors). Each sensor generates a 10Gbps stream. This leads to a total of 6,912 individual 10 Gbps streams – almost 70,000 Gbps, or 8.44 terabytes/second (TBps).

Fig. 5 outlines the data flow and processes in ASKAP. The data image cube dimensions can be varied for different observation types, as some examples will demonstrate. For continuum observations, 12,288 $n_{RA}$ x 12,288 $n_{Dec}$ x 300 $n_Z$ continuum data, where each data point holds 4 nPol x 4 byte single-precision floating-point polarizations, leads to:

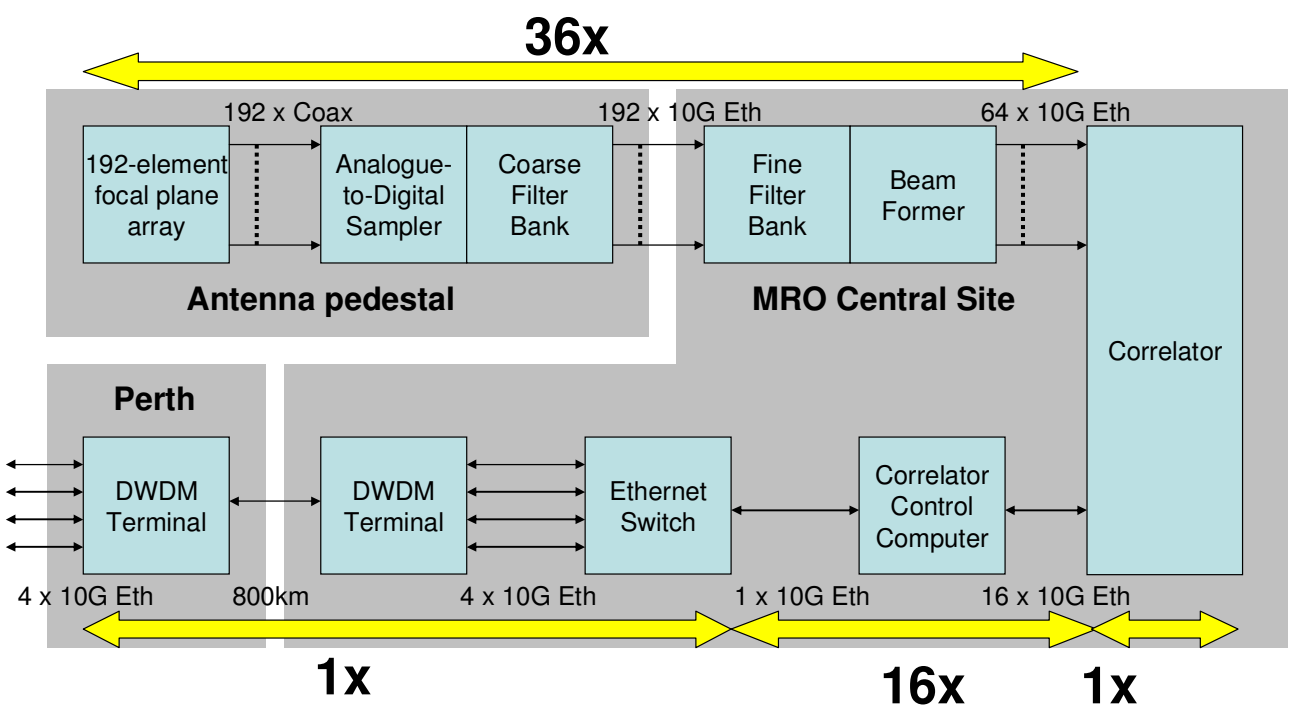

**Figure 5 – Data Flow & Processes in ASKAP**

12,288 x 12,288 x 300 x 16 bytes = 724.8 Gigabyte cube.

For spectral data observations, 4,096 $n_{RA}$ x 4,096 $n_{Dec}$ x 16,384 $n_Z$, with each data point holding 1 $n_{Pol}$ x 4 byte single-precision floating-point polarizations, gives:

4,096 x 4,096 x 16,384 x 4 bytes = 1.1 Terabyte cube.

(The array should also be capable of observing at a higher angular resolution with increased $n_{RA}$ and $n_{Dec}$.) Planned observations for ASKAP include:

- The Widefield ASKAP L-Band Legacy All-Sky Blind Survey (WALLABY) [17], generating 1200 cubes to hold a total of 1.32 Petabytes of content.
- The Deep Investigations of Neutral Gas Origins (DINGO) surveys, which will generate 960 cubes to hold a total of 1.06 Petabytes of content. DINGOb is a major SKA pathfinder experiment [18].

*Square Kilometre Array*

SKA will be a hybrid telescope, comprising a mix of technologies including single-pixel feeds, sparse aperture arrays, dense aperture arrays and phased-array feed sensors. Sizes of final data products for individual observation sets in data cubes are expected to range from 30 Terabytes up to 360 Terabytes each, with total sensor data rates generating those processed cubes varying from 0.055 Terabits/s (Tbps) up to 429 Tbps [19].

## 6. RELATED WORK

There is recognition that TCP, with its assumption that any loss due to errors is congestion requiring a decrease in throughput, does not meet the needs of radio astronomy [20] and that UDP is suitable [21]. Other UDP-based protocols, also adding acknowledgements to UDP for reliability, have been investigated for astronomy data delivery [22].

## 7. CONCLUSIONS

Radio astronomy projects pose some advanced and challenging computer networking requirements.

Our experience gained with *Saratoga*, in the analogous domain of delivering raw imagery from remote-sensing satellites, suggests that *Saratoga* will be well-suited to handling high-speed data transfer across private radio astronomy networks, allowing commercial Internet and optical Ethernet networking technologies to be leveraged by these projects.

However, just as transferring remote-sensing images to ground is only a single piece of the engineering processes that provide us with useful information about areas of the Earth, delivering astronomy data with *Saratoga* is just one small part of the vastly larger and more complex sensor and processing chain that is needed to tell us more about our universe.

## ACKNOWLEDGEMENTS

We thank Ben Humphreys of CSIRO Astronomy and Space Science (CASS) for his helpful comments on drafts of this paper.

## BIOGRAPHIES

**Lloyd Wood** is a Chartered Engineer with experience in computing, networking and aerospace. During nine years with Cisco Systems, Lloyd had responsibility for CLEO, the Cisco router in Low Earth Orbit, while working as a space initiatives manager and technical leader. With colleagues at NASA Glenn Research Center and Surrey Satellite Technology Ltd, Lloyd achieved the first tests from space of IPv6 and of the delay-tolerant networking bundle protocol for the "Interplanetary Internet." Lloyd gained his PhD from the Centre for Communication Systems Research at the University of Surrey, to where he has returned as a research fellow. He is a fellow of the Royal Astronomical Society.

**Charles Smith** is a networking consulting engineer who is currently attached to the CSIRO Astronomy and Space Science Division. He is advising the Australian Telescope National Facility and the Square Kilometre Array Project Development Offices on networking technologies, and researching architectures for the data and control planes for the ASKAP and SKA radio telescopes. Previously, Charles spent eight years at Cisco Systems, where he was a principal architect and engineered research and education networks, including the National Lambda Rail in the United States and TWAREN backbones in Taiwan.

**Wesley M. Eddy** is a network systems engineer with MTI Systems, working on projects at NASA's Glenn Research Center (GRC) and Goddard Space Flight Center (GSFC). He is co-chair of the IETF TCP Maintenance and Minor Extensions (TCPM) Working Group and of the IRTF's Internet Congestion Control Research Group (ICCRG). He holds an MS degree in Computer Science from Ohio University.

**William D. Ivancic** is a senior research engineer at NASA's Glenn Research Center, where he directs research into hybrid satellite/terrestrial networking, space-based Internet, and aeronautical Internet. Will is leading a research effort to deploy commercial off-the-shelf (COTS) technology into NASA missions, including the Space Exploration Initiative, the Airspace Systems Program and the Aviation Safety Program. Will holds BS and MS degrees in electrical engineering.

**Chris Jackson** is a Principal Engineer at Surrey Satellite Technology Ltd, where he has worked for over fifteen years, and where he designed the original specification of *Saratoga*. He has been involved with flight and ground software systems, space-to-ground protocols and flight operations for over thirty space missions.

## APPENDIX A – LINK UTILIZATION

What capacity is available for use on an optical fibre carrying a 10Gbps Ethernet link to carry data from radio astronomy sensors? A worked example, based on a simplified sensor scenario as a starting point, is given here.

The sampling throughput bitrate calculated in this trivial example is less than the available payload bitrate that the Ethernet link can support, so this rate can easily be carried and the sensor bank can be supported.

| | | |
|---|---|---|
| **Overall Ethernet capacity** | | |
| Ethernet rate | 10<br>$1x10^{10}$ | Gbps, *or*<br>bits/second |
| Ethernet jumbo frame payload size, selected for compatibility with other media (< 9000 bytes) | 8192 | bytes |
| **Ethernet frame overhead** | | |
| Minimum interframe gap | 96 | bits |
| Preamble length | 64 | bits |
| MAC source address | 48 | bits |
| MAC destination address | 48 | bits |
| Ethertype field | 16 | bits |
| Trailing CRC32c | 32 | bits |
| | **304** | **bits/frame** |
| Eth. payload size in bits | 65536 | bits/frame |
| Total frame length in bits | 65840 | bits/frame |
| **No. of frames in 10 Gbps** | **151,883.354** | **frames/second** |
| **IP transport overhead** | | |
| IPv6 header | 40 | bytes |
| UDP header | 8 | bytes |
| *Saratoga* header | 32 | bytes |
| | **80** | **bytes/frame** |
| | **640** | **bits/frame** |
| **Remaining available payload size** | **8112**<br>**64,896** | **bytes *or***<br>**bits** |
| **Maximum payload rate** | **$9.8566x10^9$** | **bits/second** |
| **Payload link utilization** | **98.566** | **%** |
| **Astronomy strawman scenario sample needs** | | |
| Sample size required | 13 | bits |
| After Nyquist doubling | 26 | bits |
| Margin needed for sample identification *etc.* – internal header overhead is spread across each sample | 4 | bits |
| | **30** | **bits/sample** |
| Total 1MHz channels or Analogue-to-Digital Converters (ADCs) | 200 | |
| Total ADC rate | $2x10^8$ | Hz |
| Oversampling factor for statistical confidence | 32/27 | |
| Oversampling rate across all ADCs | $2.37x10^8$ | Hz |
| Bitrate across all ADCs | **$7.11x10^9$** | **bits/second** |

In a more detailed and complex calculation for the more realistic ASKAP scenario, overheads including various margins, codecs and accommodations to extend the fibre distance over which Ethernet can be carried (using technology such as XAUI Attachment Unit Interfaces and XGMII Extenders, with 8b/10b encoding) must also be considered and included.

Careful engineering of the overall design would optimize the use of the available payload throughput rate supported by the link, without ever exceeding it.